\documentclass[]{article}

\usepackage[a4paper, total={6in, 9in}]{geometry}
\usepackage[title]{appendix}

\usepackage{authblk}

\usepackage[dvipsnames,table,xcdraw]{xcolor}
\usepackage{mdframed}
\usepackage{multicol}

\usepackage{cancel}
\usepackage{multirow}
\usepackage{float}
 
\usepackage{perpage} 
\MakePerPage{footnote} 
\usepackage[normalem]{ulem}

\usepackage{breqn}
\usepackage{amsmath}
\usepackage{amssymb}
\usepackage{amsfonts}
\usepackage{bm}
\usepackage{dsfont}
\usepackage{mathtools}
\usepackage{stackengine}

\stackMath
\DeclareMathOperator{\E}{\mathbb{E}}
\DeclareMathOperator{\Var}{\mathrm{Var}}
\DeclareMathOperator{\Cov}{\mathrm{Cov}}

\DeclareMathOperator*{\argmin}{arg\,min}
\newcommand*{\tran}{T}

\usepackage{booktabs}
\usepackage{multirow}

\definecolor{lightgray}{gray}{0.9}
\definecolor{background_blue}{HTML}{dde7f5}
\definecolor{figure_background_blue}{HTML}{b7c8ea}

\usepackage{tikz}
\usetikzlibrary{arrows}
\usepackage{tikzscale}
\usepackage{pgfplots}
\usepackage{subfigure}
\usepackage{graphicx}

\usepackage{hyperref,apacite}

\newtheorem{theorem}{Theorem}

\begin{document}

\title{Statistical arbitrage portfolio construction \\based on preference relations\vspace{20pt}}

\author[]{Fredi Šarić}
\author[]{Stjepan Begušić\thanks{Corresponding author. E-mail address: stjepan.begusic@fer.hr.}}
\author[]{Andro Merćep}
\author[]{Zvonko Kostanjčar}

\affil[]{\normalsize University of Zagreb, Faculty of Electrical Engineering and Computing,\\ Laboratory for Financial and Risk Analytics (\url{lafra.fer.hr}),\\ Unska 3, 10000 Zagreb, Croatia}


\maketitle

\begin{abstract}
Statistical arbitrage methods identify mispricings in securities with the goal of building portfolios which are weakly correlated with the market.
In pairs trading, an arbitrage opportunity is identified by observing relative price movements between a pair of two securities.  
By simultaneously observing multiple pairs, one can exploit different arbitrage opportunities and increase the performance of such methods.
However, the use of a large number of pairs is difficult due to the increased probability of contradictory trade signals among different pairs.
In this paper, we propose a novel portfolio construction method based on preference relation graphs, which can reconcile contradictory pairs trading signals across multiple security pairs. The proposed approach enables joint exploitation of arbitrage opportunities among a large number of securities. Experimental results using three decades of historical returns of roughly 500 stocks from the S\&P 500 index show that the portfolios based on preference relations exhibit robust returns even with high transaction costs, and that their performance improves with the number of securities considered.
\end{abstract}

Keywords: statistical arbitrage, pairs trading, preference relations, potential method.

\section{Introduction}
\begin{figure}[h]
\begin{center}
\includegraphics[width=0.75\linewidth]{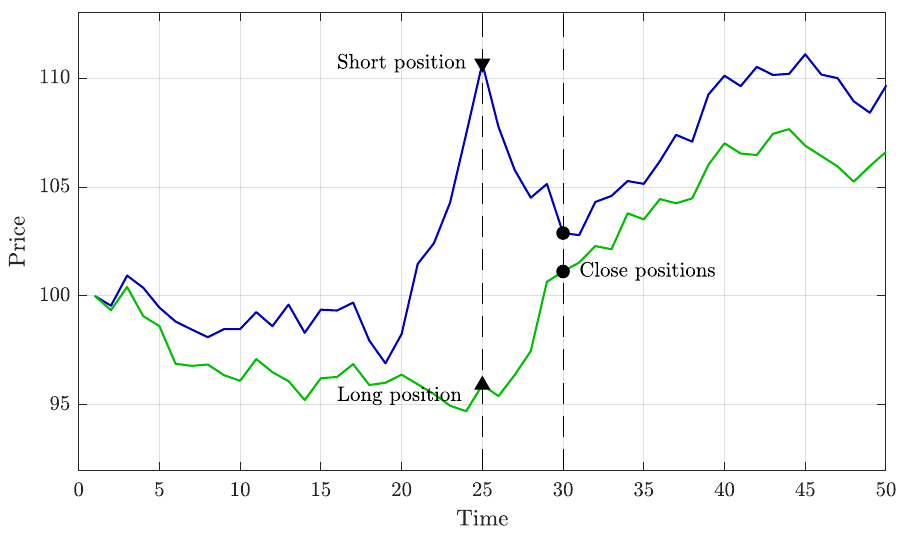}
\caption{Principles of pairs trading, shown through an example of two simulated price trajectories. At the point of detected mispricing, the underpriced security is bought (long position) and the overpriced is sold (short position). Profit is taken (positions closed) if the market is able to correct the mispricing.}
\label{fig:pairs_trading_example}
\end{center}
\end{figure}

Signal processing methods are becoming increasingly important in financial applications, providing added value in risk management \cite{Johnston2011}, time series modelling \cite{Zhang2017} and extracting knowledge from complex market systems \cite{Kostanjcar2016}. A notable application lies in finding profitable trading strategies based on time series methods, such as the \emph{statistical arbitrage} family of strategies, which exploit the abrupt short term overpricing and underpricing of financial securities \cite{Elliott2005,Chen2019}. Such situations are identified by examining relative pricing between securities rather than by estimating an absolute value of the security. If two securities have similar characteristics, then the prices of those securities should behave similarly. Therefore, if an investor is able to find two securities with similar characteristics, their price combination could be modeled as a stationary process with a constant mean \cite{Krauss2017}. A significant deviation from the mean indicates mispricing between two securities, where one security is underpriced, other overpriced, or the mispricing is a combination of both. In such situations, an investor would buy the underpriced security and short sell the overpriced. An investor would profit when the mispricing between securities is corrected. The ability of the market to correct these mispricings is one of the fundamental assumptions of pairs trading methods. The principles of such strategies (also known as pairs trading) are shown in Figure \ref{fig:pairs_trading_example}. For a more in-depth review of pairs trading strategies, see \cite{Krauss2017}. 

Statistical arbitrage techniques have been utilized in the financial industry for a long time \cite{Vidyamurthy2004,Elliott2005}, but real interest within the academic community has been observed with the publication of the seminal paper by Gatev et al. \cite{Gatev2006}. They use two distinguishable time windows: (i) $12$ months pairs formation and (ii) $6$ months trading period. All pairs of securities are ranked according to the sum of squared differences (SSD) of the normalized price series on a formation period. Top pairs with the minimal value of SSD are then observed in the next $6$ months - during the trading period. A simple distance based strategy is proposed: once the log-price spread for a pair of securities deviates significantly from its mean, a long position is entered in the undervalued security, and short in the overvalued security at equal proportions, forming a self-financing two-asset portfolio. These positions are held until the price spread reverts back to its mean or until the end of the trading period. This principle has been employed in a number of approaches 
focused on novel methods for idenitying arbitrage opportunities and selecting pairs \cite{Huck2009,Ramos-Requena2017}.

However, recent studies on the profitability of pairs trading strategies \cite{Do2010,Rad2016} have shown a decline in the profitability of simple pairs trading strategies in developed markets. The lack of profitability is mainly attributed to two factors: (i) inability of choosing convergent pairs in the presence of large number of securities; (ii) the increased use of pairs trading strategies, i.e. less mispricing opportunities. 
Subsequent surge in research interest regarding statistical arbitrage and pairs trading revolved around improving the pairs selection quality, determination of optimal trade enter and exit times, and balancing statistical arbitrage portfolios \cite{Zhao2019,Ramos-Requena2020}. 
In a follow-up work \cite{Huck2015} another study of pairs trading methods was performed using S\&P500 constituents from 2000 to 2011. Their findings support the hypothesis that the distance-based method \cite{Gatev2006} is unable to robustly select convergent pairs. However, other approaches based on cointegration do exhibit high and robust positive returns and are able to significantly reduce risk of a pair spread not reverting to its long term mean \cite{Huck2015a,Rad2016,Clegg2018}.

Jacobs et al. \cite{Jacobs2015} identify two possible factors that drive profit of pairs trading strategies: (i)  overreaction of investors to the firm-specific shocks (`high attention hypothesis'), and (ii) slow diffusion of common market shock information into subset of securities (`limited attention hypothesis'). Their findings are more supportive of limited attention hypothesis. This is especially true in markets with a large number of tradeable financial instruments, due to the `information overload' which leads to limited visibility of individual pairs, thus reducing common shock information diffusion even further. The limited attention hypothesis provides empirical motivation for \emph{multivariate pairs trading} and statistical arbitrage techniques  \cite{Avellaneda2010,Krauss2017a,Mrcela2017,Zhao2018,Zhao2019} which consider collections of securities rather than individual pairs.

To tackle the multivariate pairs trading problem, Huck \cite{Huck2009,Huck2010} applies a machine learning and combined forecasting approach.
The method relies on future return predictions, implemented by neural network models, to obtain expected spreads. The ELECTRE-III method by Hashemi et al. \cite{Hashemi2016} is used to rank all securities according to the deviations from predicted spreads. 
Final portfolio is constructed using total ordering obtained from the ranking method. Top $p$ are securities are selected for a long position and bottom $p$ securities for a short position. Long and short positions are entered at equal proportions, forming a dollar neutral portfolio. 
A caveat of the proposed method is its reliance on the reliable future return predictions, which are very hard to obtain \cite{Timmermann2004}. 

Recently, Zhao et al. \cite{Zhao2018,Zhao2019} proposed a method for optimizing the portfolio of spreads under the multivariate cointegration assumption, seeking those with the strongest mean-reverting behavior and higher variance (in order to increase the profit potential). However, in the multivariate cointegration framework the price (or log-price) series are assumed to have a stationary linear relationship - an assumption that yields elegant mathematical properties but only allows for specific multivariate co-movement structures.

In this paper we allow for the use of an arbitrary pair formation method and focus on reconciling the resulting contradictory signals \cite{Mrcela2017}. We employ a \emph{multiple-criteria decision-making} (MCDM) technique, namely the \emph{potential method} \cite{Caklovic2012}, to produce meaningful ranking of securities, even if the pairwise information between two securities is contradictory \cite{Mrcela2017}. The potential method scales well with the number of considered security pairs, 
enabling the use of extremely large number of securities. 
A portfolio construction technique is proposed, based on a more informative partial ordering, rather than total ordering. We also develop a novel method for reducing portfolio turnover and volatility. Extensive assessments of developed techniques are preformed, and the results show that the proposed method generates robust portfolios that are weakly correlated with the market and are profitable even in the regime of high transaction costs. In addition, we provide a theoretical motivation and empirical evidence that performance of our method generally improves with the number of considered securities, which is in line with the limited attention hypothesis. 

It is important to note that the main focus of the paper is on the proposed preference relation based decision making mechanism and its application in reconciling contradictory signals in a given trading application, together with the set of tools that allow these applications. To this end, the pairs trading rule considered in the experimental setup is purposely straightforward and simple. Hopefully, the proposed approach will motivate future applications with other more sophisticated trading rules and applications.

The rest of this paper is organized as follows. Firstly, in section \ref{sec:methodology}, the details of the proposed framework are given, including the preference relation estimation, preference graph transformations, and portfolio construction methods.
In section \ref{sec:experimental_setup} we describe the experimental framework setup and the data used to obtain results. 
Section \ref{sec:results} presents the evaluation results of our method on more than $500$ U.S. stocks in the period from 1990 to 2019. Finally, in section \ref{sec:conclusion} we discuss the caveats of proposed method, propose directions of further research, and give concluding remarks.

\section{Methodology}\label{sec:methodology}

Let a binary relation $M: \mathcal{S}\times\mathcal{S} \xrightarrow{} \{-1, 0, 1\}$ represent a pairs trading strategy, where $\mathcal{S}=\{s_1, s_2, \ldots, s_N\}$ is a set of tradeable securities.\footnote{Formally, a binary relation $M$ is a subset of the Cartesian product $\mathcal{S}\times\mathcal{S}$, and consists of elements $(s_i, s_j) \in M$. Here we abuse the notation a little in order to bridge a gap between binary relations and pairs trading, in order to express indifference. Situations where $(s_i, s_j)\not\in M \land (s_j, s_i)\not\in M$ are represented as $M(s_i, s_j) = M(s_j, s_i) = 0$.} When $M(s_1, s_2) = 1$ a long position should be entered in the first security $s_1$ and a short position in the second security $s_2$, when $M(s_1, s_2) = -1$ a short position should be taken in $s_1$ and long position in $s_2$, and in the case $M(s_1, s_2) = 0$ neither $s_1$ nor $s_2$ should be bought or sold. However, due to the stochastic and non-stationary characteristics of the market, $M$ can produce contradictory outputs.
For example, if $M(s_1, s_2) = 1$, $M(s_2, s_3) = 1$, and $M(s_1, s_3) = -1$ a long/short portfolio could not be formed without violating some of the outputs of $M$. Fortunately, contradictory outputs can be avoided if $M$ is constrained to be a \emph{preference relation}. A preference relation is a binary relation with following properties:
\begin{enumerate}
    \item \textit{Irreflexivity}: $\forall\, s\in \mathcal{S}: \; M(s, s) = 0$ (a security $s$ can not be preferred over itself),
    \item \textit{Asymmetricity}: $\forall\, s_i, s_j\in \mathcal{S}: \; \left(M(s_i, s_j) = 1\right) \implies \left(M(s_j, s_i) = -1\right)$ (if $s_i$ is preferred over $s_j$, then $s_j$ can not be preferred over $s_i$),
    \item \textit{Transitivity}: $\forall\, s_i,s_j,s_k\in \mathcal{S}: \; \left(M(s_i, s_j)=1\right) \; \land \; \left( M(s_j, s_k) = 1\right) \implies \left( M(s_i, s_k) = 1\right)$ (if $s_i$ is preferred over $s_j$ and $s_j$ is preferred over $s_k$, then $s_i$ must also be preferred over $s_k$).
\end{enumerate}

\noindent These conditions induce \emph{strict partial ordering} over a set of securities $\mathcal{S}$, allowing an investor to rank securities according to some criterion, and ultimately form a portfolio which may include long positions for the most highly ranked securities and short for the lowest ranked ones.
The problem of constraining $M$ to be a preference relation can be relaxed into finding a preference relation $M^*$ which is most similar (w.r.t. some metric) to the binary relation $M$:
\begin{align}\label{eq:core_problem}
    M^* = \argmin_{M'\in\mathcal{M}^*} d(M, M')
\end{align}
where $M'$ is some preference relation, $d:\mathcal{M}\times\mathcal{M}\xrightarrow{} \mathbb{R}^+$ is a distance metric,  $\mathcal{M}$ is a space of binary relations, and $\mathcal{M}^*\subset\mathcal{M}$ is a space of preference relations.

\subsection{Estimating preference relations}
 This amounts to the problem of finding a preference relation $M^*$ closest to a binary relation $M$ (which does not necessarily satisfy the preference relation properties), which is a generally topic within the field of multiple-criteria decision making (MCDM).
MCDM methods generally perform the optimization over a real-valued generalization of the preference relation, namely a \emph{preference function} $\rho^*$, since it is often more natural to represent a preference between two alternatives on a real-valued scale \cite{Mrcela2017}. 

The potential method defines a preference function $\rho^*: \mathcal{S}\times\mathcal{S}: \xrightarrow{}\mathbb{R}$ as the one for which there exists a latent utility function $u^*$ such that the following condition is satisfied:
\begin{align}\label{eq:pm_pref_func_cond}
    \rho^*(s_i, s_j) = u^*(s_i) - u^*(s_j).
\end{align}
A utility function $u^*: \mathcal{S}\xrightarrow{}\mathbb{R}$ expresses total preference of some security $s_i$, and once it is inferred it can be used to rank securities, that is it produces a total order among securities. Basically, if the condition \eqref{eq:pm_pref_func_cond} is satisfied, the preference of $s_i$ over $s_j$ is defined by the difference in their utilities. Assuming Eq. \eqref{eq:pm_pref_func_cond} holds, additional properties of preference function $\rho^*$ can be derived:
\begin{align}
    &\rho^*(s_i, s_i) = u^*(s_i) - u^*(s_i) = 0,\quad \label{eq:irreflexivity_pm_property}\\
    &\rho^*(s_i, s_j) = -\rho^*(s_j, s_i),\quad \label{eq:asymmetricity_condition}\\
    &\left(\rho^*(s_i, s_j) > 0\right) \land \left(\rho^*(s_j, s_k) > 0 \right)\implies \left(\rho^*(s_i, s_k) > 0\right). \quad \label{eq:transitivty_pm_property}
\end{align} 
These are namely the \textit{irreflexivity}, \eqref{eq:irreflexivity_pm_property}, \textit{assymetricity} \eqref{eq:asymmetricity_condition} and \textit{transitivity} \eqref{eq:transitivty_pm_property} properties, all of which arise directly from Eq. \eqref{eq:pm_pref_func_cond}, and correspond to the ones introduced in the first paragraph of Section \ref{sec:methodology}. The preference relation $M^*$ can then be simply recovered from preference function $\rho^*$, for instance by taking the sign of the preference function.
A function $\rho^*$ needs to satisfy condition \eqref{eq:pm_pref_func_cond} for all security pairs $(s_i, s_j)\in\mathcal{S}\times\mathcal{S}$ in order to be considered a preference function. Therefore, it is beneficial to express condition \eqref{eq:pm_pref_func_cond} for all security pairs at once by writing it in a matrix form as:

\begin{align}\label{eq:potential_method_condition}
    \bm{Bu}^* = \bm{\rho}^*
\end{align}
where $\bm{B}$ is a matrix containing elements $B_{i,j}\in\{-1, 0, 1\}$ and has the form of a directed graph incidence matrix, $\bm{\rho}^* = [\rho^*(s_1, s_1), \rho^*(s_1, s_2), \ldots, \rho^*(s_N, s_N)]$ is vector of preference function $\rho^*$ evaluations for all security pairs, and $\bm{u}^*=[u^*(s_1), u^*(s_2), \ldots, u^*(s_N)]$ is vector of security utilities\footnote{In \cite{Caklovic2012} the vector of utilities is referred to as the vector of node potentials, hence the name potential method.}. Since the matrix $\bm{B}$ has the structure of incidence matrix, the Equation $\eqref{eq:potential_method_condition}$ represents a graph $\mathcal{G}^*$, named the \emph{preference graph}. The preference graph $\mathcal{G}^*$ is a complete weighted directed graph
in which each vertex represents a security and each edge represents the preference function $\rho^*$. The sign of the  preference function is encoded as an edge direction, and the intensity of the preference is represented as an edge weight. 
 

Preference function $\rho^*$ can be easily characterized using techniques employed by most statistical arbitrage methods. 
Most statistical arbitrage methods compute the deviation of the security pair price spread from its long term mean in order to measure degree of arbitrage opportunity, and use it to perform trading actions. The deviation of the price spread can be used as a preference function $\rho^*$. However, in general, most statistical arbitrage methods do not provide any guarantee with regards to transitivity property \eqref{eq:transitivty_pm_property}, and as such will not always satisfy potential method condition \eqref{eq:potential_method_condition} for any $\bm{u}^*$. This problem can be solved via the potential method by finding the best approximation of the function $\rho$ which satisfies the preference function properties (\ref{eq:irreflexivity_pm_property}, \ref{eq:asymmetricity_condition}, \ref{eq:transitivty_pm_property}). More formally, let $\rho: \mathcal{S}\times\mathcal{S}\xrightarrow{}\mathbb{R}$ be a statistical arbitrage method, i.e. the deviation of the price spread, which \emph{does not satisfy the condition \eqref{eq:potential_method_condition} for any $\bm{u}^*$}, then one can obtain a preference function vector $\bm{\rho}^*$ by solving an optimization problem:
\begin{align}\label{eq:potential_method_optimization}
    \bm{\rho}^* =& \argmin_{\bm{\rho}'\in\mathcal{P}^*} \quad d(\bm{\rho}', \bm{\rho})\\
    \textrm{s.t.}     \quad & \bm{B}\bm{u}^* = \bm{\rho}^* \nonumber \\
    & \sum_{i=0}^N u_i^* = 0   \nonumber
\end{align}
\noindent where $\mathcal{P}^*$ is the set of all preference function vectors $\bm{\rho}^*$, $\bm{\rho}$ a vector of evaluations of $\rho$ for all security pairs, and $d$ is some distance function. The constraint $\sum_{i}^N u_i^* = 0$ is needed in order to obtain a unique solution, because the preference relation obtained from a utility function is invariant to scaling, i.e. $s_i \succ s_j = \alpha u^*(x) > \alpha u^*(y)\quad\forall {s_i,s_j}\in\Omega,\,\forall\alpha\in\mathbb{R}^+/\{0\}\ $. By choosing the Euclidean distance as the distance function $d$, the optimization problem \eqref{eq:potential_method_optimization} has an analytical solution:
\begin{align}
    \bm{u}^* &= \frac{1}{N}\bm{B}^\tran\bm{\rho} \label{eq:potential_method_utility_estimation}\\
    \bm{\rho}^* &= \bm{Bu}^*
\end{align}
\noindent The derivation and implementation details of the potential method are presented in the appendix section (\ref{sec:apx:potential_method_derivation}). Additionally, under some assumptions, the potential method is an unbiased and consistent estimator of utility $\bm{u}^*$ and preference function $\bm{\rho}^*$ as stated in Theorem \ref{theorem:pm_estimator}.

\begin{theorem}\label{theorem:pm_estimator}
Let the $\bm{\rho}$ be a vector of $\rho$ function evaluations for all security pairs, which may not satisfy potential method condition \eqref{eq:potential_method_condition} for any utility vector $\bm{u}^*$, expressed as:
\begin{align}
    \bm{\rho} = \bm{\rho}^* + \bm{\epsilon}
\end{align}
\noindent where $\bm{\rho}^*$ is the vector of evaluations of the preference function $\rho^*$, and $\bm{\epsilon}$ is some additive zero-mean noise. The potential method is an unbiased estimator of utility vector $\bm{u}^*$ and preference function vector $\bm{\rho}^*$. In addition, if the noise terms are uncorrelated across different pairs, then the variance of the potential method estimator diminishes with the increase in the number of considered securities. Thus, the potential method is also a consistent estimator of $\bm{u}^*$ and $\bm{\rho}^*$.
\end{theorem}

\noindent The proof of Theorem \eqref{theorem:pm_estimator} is provided in the Appendix \ref{sec:apx:potential_method_estimator}. Theorem \eqref{theorem:pm_estimator} provides theoretical motivation for a use of the potential method as a technique for finding a preference function $\rho^*$ in the presence of a large number of considered securities.

\subsection{Preference preserving transformations}
Finally, once the preference function vector $\bm{\rho}^*$ is calculated, we need to compute a preference relation $M^*$. This can easily be done by taking the sign of the preference function. However we argue that only taking the sign of the preference function $\bm{\rho}^*$ on the raw outputs of the potential method can lead to undesirable effect of a selection of non-arbitrage positions.  
This problem can be mitigated by transforming a preference function $\rho^*$ via a \emph{preference preserving transformation} before constructing the preference relation $M^*$, by taking the sign of the transformed preference function. Preference preserving transformations are transformations of the preference function which preserve preference function properties (\ref{eq:irreflexivity_pm_property},\ref{eq:asymmetricity_condition},\ref{eq:transitivty_pm_property}).

\subsubsection{Edge thresholding}

Firstly, we can improve the selection of arbitrage positions by noticing that the values of the preference function $\rho^*$ will rarely be equal to zero, since they represent a deviation of the security price spread. Consequently, such statistical arbitrage method  will almost always suggest long or short position for every security $s\in\mathcal{S}$.  
A common way of dealing with this issue is to use a thresholding rule. Thresholding transformation of the preference function $\rho^*$ is defined as:
\begin{align}
    \rho_{k}^*(s_i, s_j) =
    \begin{cases}
        \rho^*(s_i, s_j) &\textrm{if } |\rho^*(s_i, s_j)| \geq \kappa\\
      0&\textrm{otherwise}
    \end{cases}
\end{align}
\noindent where $\kappa$ is some threshold value. Values of the thresholding parameter $\kappa$ should be carefully selected based on preference function being used. A preference function whose outputs are standardized with mean $0$ and standard deviation of $1$ greatly simplifies the selection of thresholding parameter $\kappa$. For example, $\rho^*(s_i,s_j) \geq 3$ would signify a $3\sigma$ event, which might be a good indicator of an arbitrage opportunity. Thresholding transformation is a valid preference preserving function, as shown in appendix section (\ref{sec:preference_preserving_transformations}). 


\subsubsection{Vertex pruning}
In order to further improve selection of arbitrage positions we propose another preference preserving transformation which exploits graph structure of preference function.
A preference relation $M^*$ induces a strict partial ordering among securities, and this ordering is preserved in the $\rho^*$ and can be represented as a  complete directed acyclic graph $\mathcal{G}^*$. 
Directed acyclic graphs by definition have some vertices which have a zero in-degree $d^{(in)}$ and some vertices which have a zero out-degree $d^{(out)}$. Vertices which have $d^{(in)}(s_i)=0$ correspond to the securities which are preferred among all other securities and are good candidates for a long position, i.e. if $\left(d^{(in)}(s_i) = 0\right) \land \left(d^{(out)}(s_i) > 0\right) \iff \left(\rho^*(s_i, s_j) \geq 0\right)\quad\forall i \neq j$. Conversely, a vertex with out-degree equal to $0$ corresponds to a security which is least preferred among all securities and is good candidate for a short sell position, i.e. $\left(d^{(out)}(s_i) = 0\right) \land \left(d^{(in)}(s_i) > 0\right) \iff \left(\rho^*(s_i, s_j) \leq 0\right)\quad\forall i \neq j$. All securities on the path between most preferred and least preferred securities are not considered to be good candidate for a trading position since stronger arbitrage signal exists. We therefore remove intermediate vertices on every path from $0$ in-degree vertices to $0$ out-degree vertices as:
\begin{align}
    \rho_p^*(s_i, s_j) = \begin{cases}
      \rho^*(s_i, s_j) &\text{if } \left(d^{(in)}(s_i) = 0\right)\ ||\ \left(d^{(out)}(s_i) = 0\right), \\
      0 &\text{otherwise}.
    \end{cases}
\end{align}
The removal of intermediate vertices in a preference graph is also a preference preserving transformation, as shown in appendix section \ref{sec:preference_preserving_transformations}. 

\subsubsection{Final vertex selection}
Lastly, we propose a third preference preserving transformation which utilizes a utility $u$ information provided by the potential method in principled manner. Only top $n$ and bottom $m$ vertices (i.e. securities), w.r.t. the utility $u^*$, in the preference graph $G^*$ are preserved, and all other vertices and their corresponding edges are removed. By setting $m=0$ the preference relation $M^*$ will only produce long trading signals, and conversely setting $n=0$ will produce only short sell trading signals. This transformation is complementary to the thresholding transformation which removes edges in preference graph, while this transformation removes vertices of the preference graph. Keeping only top and bottom securities in preference graph generally is not a preference preserving transformation. However, when the preference graph has the structure of the bipartite graph, which is ensured with the removal of intermediate vertices, then this transformation will keep such structure and in that regard this is also preference preserving transformation. The proof of the preference preserving property is given in appendix section \ref{sec:preference_preserving_transformations}.


Finally, the preference relation $M^*$ is computed by firstly applying thresholding transformation, then removing intermediate vertices and lastly selecting top $n$ and bottom $m$ vertices w.r.t. utility $u^*$ in that order, and then taking the sign of the transformed preference function.





\subsection{Asset allocation}
Finally, after deciding which securities to take a long position in, and which to short sell, an investor needs to decide how to allocate the weights in a portfolio of selected assets. The aim of the portfolio allocation step is to balance the risk to profit trade-off. 
We suggest a novel, utility proportional portfolio allocation scheme, which is tailored for use in the statistical arbitrage context. Let $u^*(s_i)$ be the utility of security $s_i$, calculated using the potential method. Then, the weight $w(s_i)$ of security $s_i$ is calculated as:

\begin{align}
    w(s_i) = 
    \begin{cases}
      \phantom{-}\dfrac{|u^*(s_i)|}{\sum_{s_j \in T_L} |u^*(s_j)|}  & \text{if } s_i \in T_L, \\[12pt]
      -\dfrac{|u^*(s_i)|}{\sum_{s_j \in T_S} |u^*(s_j)|}  & \text{if } s_i \in T_S, \\[12pt]
      \phantom{-}\, 0 & \text{otherwise},
    \end{cases}  
    \label{portfolio_weights}
\end{align}
where $T_L$ is the set of securities in the long leg of the portfolio. Conversely $T_S$ is a set of securities in the short leg of the portfolio. This kind of wealth allocation schema will result in zero investment portfolios when both short and long positions are allowed. 

\subsection{Reducing turnover}
Statistical arbitrage methods exploit short term events of market inefficiencies, resulting in large investment shifts, thus high turnover is expected. Strategies produced by combining aforementioned methods may exhibit very short holding periods and high turnover. In addition to inflated turnover, in the presence of the short holding periods many positions may be closed before the expected profits were realised, thus lowering the profitability of the trading strategy even more. The reason behind this phenomena is mostly due to how long and short positions are chosen - the proposed conditions for entering a long or short position do not use information about the current positions of securities included in the portfolio. Therefore, potentially profitable securities may be removed from the portfolio in the presence of seemingly better opportunities, even if the profit from the position has not yet been realised. In order to mitigate this issue, a security position is held active (long or short) until it is determined that it no longer seems profitable or until some profits are made and more profitable position is detected. This can be done using the information in the preference relation graph of the considered securities. A position is considered profitable until the sign of the utility of that security is reversed or becomes $u^*(s_i) = 0$. We term this technique the \textit{momentum decorator}\footnote{Securities that are selected to be included in portfolio are harder to take out of portfolio, hence they have `momentum'. The `decorator' part of the name comes from the fact that this method is general enough and can be applied to any method in the presence of utility $u$, hence `decorating' it.}.

\section{Experimental setup}\label{sec:experimental_setup}
In order to assess the quality and robustness of the proposed approach we perform extensive backtesting using daily prices of a total of $N = 553$ U.S. stocks in the period from 1990 to 2019. 
In order to test the robustness of the proposed method with respect to the security subset selection, we propose a \emph{security bootstrapping} approach. First, a bootstrap set of securities is selected as a random subset of securities of fixed size $N' < N$ from the original (full) set of securities. The considered strategy is backtested on the bootstrap set and its performance is evaluated. The process is repeated a number of times $B$ (subsets of securities are allowed to overlap), with the same fixed parameter configuration. By doing so, the robustness of the trading strategy to the selected security subset can be examined and by varying the sizes of bootstrap set, we can inspect how the number of securities affects performance. In our results, we use $B=100$ security bootstrap samples of sizes $N''=50$ and $N'=250$, and finally we evaluate strategies on the full set of $N=553$ stocks. All the backtests were run with the same set of parameters, displayed in Table \ref{tbl:parameters}. Note that a strategy using the \emph{momentum decorator} can produce more than $20$ trade signals for both long and short positions. 

\begin{table}[H]
\renewcommand{\arraystretch}{1.3}
\caption{Parameters of the tested strategies.}
\label{tbl:parameters}
\centering
\begin{tabular}{cc}\toprule%
Transaction costs & $0.1\%$\\
Lookback period & $60$ days\\
Thresholding parameter $\kappa$ & $3.0$\\
Trade signal limits ($m=n$) & $20$\\ \hline%
\end{tabular}
\end{table}

Since the portfolio formed by the proposed method is reweighted and rebalanced on a daily basis, the effect of slippage and transaction cost is significant, especially when portfolio turnover is high. In order to provide more realistic performance metrics and test the robustness of the strategy, we use higher than usual transaction costs (most of the pairs trading research report a transaction costs of $0.05\%$ \cite{Avellaneda2010}) and we use open prices of the next day to execute the market orders. To calculate total returns over multiple time periods, the entire capital is assumed to be fully invested in the given portfolio each day, with the weights as defined in Eq. \eqref{portfolio_weights}.

\subsection{Preference function}\label{sec:preference_function}

In the statistical arbitrage context, a function $\rho$ expresses the degree of an arbitrage opportunity between two securities $(s_i, s_j)$. This preference is often expressed as a deviation of the price spread at the time $t$ from the expected value of the price spread. Inspired by the work of {\cite{Gatev2006}} we utilize the following preference function:
\begin{align}
    \rho(s_i, s_j)^{(t)} =\frac{c_{i,j}^{(t)} - \mu_{i,j}^{(t)}}{\sigma_{i,j}^{(t)}}
\end{align}
\noindent where the $c_{i,j}^{(t)}$\footnote{Here we sacrifice correctness of notation for the purpose of simplicity. $x_{i,j}^{(t)}$ represents evaluation of function $x$ whose arguments are data of securities $s_i$ and $s_j$ available at the time $t$, that is $x(s_i, s_j, t)$} is the price spread between securities $s_i$ and $s_j$ at timestamp $t$, and the parameters $\mu_{i, j}^{(t)}$ and $\sigma_{i, j}^{(t)}$ are the mean and the standard deviation of the price spread $c_{i, j}^{(t)}$ respectively. The price spread between securities $s_i$ and $s_j$ at particular timestamp $t$ is defined as:
\begin{align}
    c_{i, j}^{(t)} = \log\frac{p_{i}^{(t)}}{p_{j}^{(t)}}
\end{align}
\noindent where $p_{i}^{(t)}$ and $p_{j}^{(t)}$ denote historical prices of securities $s_i$ and $s_j$ at the time $t$ respectively. This corresponds to the commonly used log-return between these two prices -- due to the logarithm the spreads are centered around zero and symmetric. In \cite{Gatev2006}, the price spread is calculated as the log ratio of normalized price series. However, we normalize the deviation from the expected price spread, not the prices. This modification ensures that the preference functions of different security pairs are comparable, alleviating the problem of choosing the thresholding parameter $\kappa$. 

The parameters $\mu_{i, j}^{(t)}$ and $\sigma_{i, j}^{(t)}$ are estimated on a lookback period $\left[t - T, t\right]$ using an unbiased estimator:
\begin{align}
    \mu_{i,j}^{(t)} &= \frac{1}{T-1} \sum_{\tau=t-T}^{t-1} c_{i,j}^{(\tau)} \label{eq:ggr_mean}\\
    \sigma_{i,j}^{(t)} &= \sqrt{
        \frac{1}{T-2}\sum_{\tau=t-T}^{t-1} \left(c_{i,j}^{(\tau)} - \mu_{i,j}^{(\tau)}\right)^2
    }\label{eq:ggr_std}
\end{align}
\noindent Notice that the summation in Equations (\ref{eq:ggr_mean}) and (\ref{eq:ggr_std}) are only up to the $t-1$ and the information about the price spread at the time step $t$ is intentionally left out, since we do not want the current value of the spread to dampen the value of preference function.

\subsection{Implementation of the strategy}
Now when all of the components of the strategy in this experimental setup are introduced, we can provide an outline for the proposed approach in the form of a meta-algorithm. At each time $t$, given the securities in the (bootstrapped) security set, the following is performed:
\begin{itemize}
    \item[(i)] Calculate the preference functions $\rho(s_i, s_j)^{(t)}$ for all pairs $(s_i, s_j)$ (note that these are not consistent as they generally contain contradictory signals and cannot directly be used to form a portfolio). 
    \item[(ii)] Calculate the consistent preference relations $\rho^*(s_i, s_j)^{(t)}$ from $\rho(s_i, s_j)^{(t)}$ and apply the preference preserving transformations of the preference relation graph. The final vertex selection provides the security selections for the long and short portfolios. 
    \item[(iii)] Construct the portfolios using the considered weight allocation scheme.
\end{itemize}
In addition to the portfolio weighting scheme proposed in Eq. \eqref{portfolio_weights}, in this experimental setup we also consider an equal-weighting scheme (all securities within the long and short portfolios are assigned equal weights), in order to gauge the effect of the portfolio weights on the performance.

\section{Results}\label{sec:results}
Firstly, to study the change in performance with respect to the number of considered securities, we examine the returns of the long-short portfolios formed by using the proposed methodology with the \emph{momentum decorator} and the utility proportional weighting scheme, over the bootstrap subsets for $N'' = 50$ and $N' = 250$ stocks, and the full set of $N =553$ stocks. Table \ref{tbl:results1} shows the backtested strategy returns in excess of the risk-free rate. Among the $B=100$ bootstrap subsets, the median result and the $95\%$ confidence intervals (the $2.5\%$ and $97.5\%$ percentiles) are displayed for the different numbers of securities. Since there is only one (full) set of securities, no confidence intervals are reported for the $N=553$ case, and the median corresponds to the single backtested full sample of stocks.

\begin{table}[h]
\renewcommand{\arraystretch}{1.3}
\caption{Statistics of annualized returns for the long-short portfolios formed on bootstrap subsets for the different numbers of securities. For each of the portfolio statistics, the median (above) and 95\% confidence intervals (in brackets below) are displayed.}
\label{tbl:results1}
\centering
\begin{tabular}{lccc}%
    \hline
    &      $N'' = 50$          & $N' = 250$             & $N =553$  \\\hline  
    \multirow{2}{*}{Mean}
    & $-0.54\%$           & $8.95\%$               & $14.03\%$                \\
    & $(-5.4\%, 7.05\%)$  & $(5.14\%, 11.49\%)$    & $-$                      \\[6pt]
    \multirow{2}{*}{Std. dev.}
    & $21.18\%$           & $17.24\%$              & $18.95\%$                \\
    & $(18.61\%, 24.76\%)$ & $(16.35\%, 18.21\%)$  & $-$                      \\[6pt]
    \multirow{2}{*}{t-Statistic}
    & $-0.14$             & $2.74$                 & $3.94$                   \\
    & $(-1.31, 1.87)$     & $(1.53, 3.52)$         & $-$                      \\\hline
\end{tabular}%
\end{table}
The results in Table \ref{tbl:results1} suggest that portfolio performance improves when the selected security subset contains more securities - this is expected and in line with limited attention hypothesis, since the smaller security subsets almost surely lack all the valid pairs of stocks for arbitrage positions. However, the portfolios are already robust to the selection of specific securities when $N' = 250$ stocks are considered - as indicated by the positive excess returns. Even though the performance of these portfolios is not as strong as the one for the full security set ($N = 553$), the statistics given by the bootstrap security subsets indicate that the proposed approach works well. This implies an important finding: the approach works well regardless of the selected security subset, as long as the size of the subset is large enough. 

\begin{figure}[H]
\begin{center}
\includegraphics[width=0.75\linewidth]{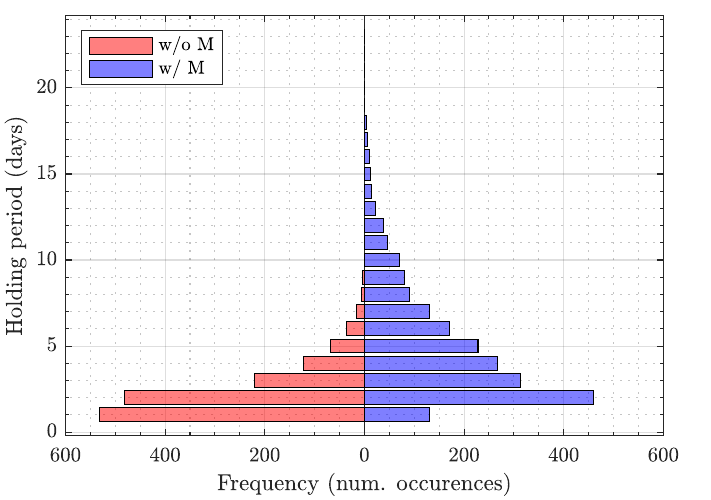}
\caption{Histograms of the holding periods for each position of the strategies with and without the momentum decorator.}
\label{momentum_decorator_hist}
\end{center}
\end{figure}

To examine the effect of the momentum decorator, we also evaluate the holding period of each position for the strategy with and without the momentum decorator (labeled w/ M and w/o M), as shown in the histograms in Figure \ref{momentum_decorator_hist}. These results demonstrate the effect of the momentum decorator method on the holding periods - instead of positions being held a few days in most cases (left histogram), the proposed method achieves a longer holding periods across of up to 23 days, which allows positions more time to be profitable and reduces turnover. 

\begin{figure}[H]
\begin{center}
\includegraphics[width=0.75\linewidth]{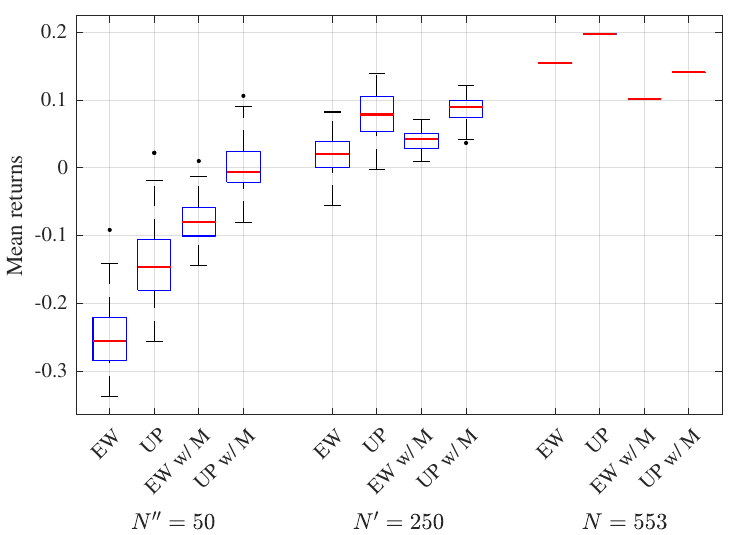}
\caption{Annualized mean returns across bootstrap security subsets for the 4 different portfolio forming methods and the selected numbers of securities. Since the $N=553$ stocks case contains the full sample, the results for this case are single points rather than boxplots.}
\label{boxplot1}
\end{center}
\end{figure}

\begin{figure}[H]
\begin{center}
\includegraphics[width=0.75\linewidth]{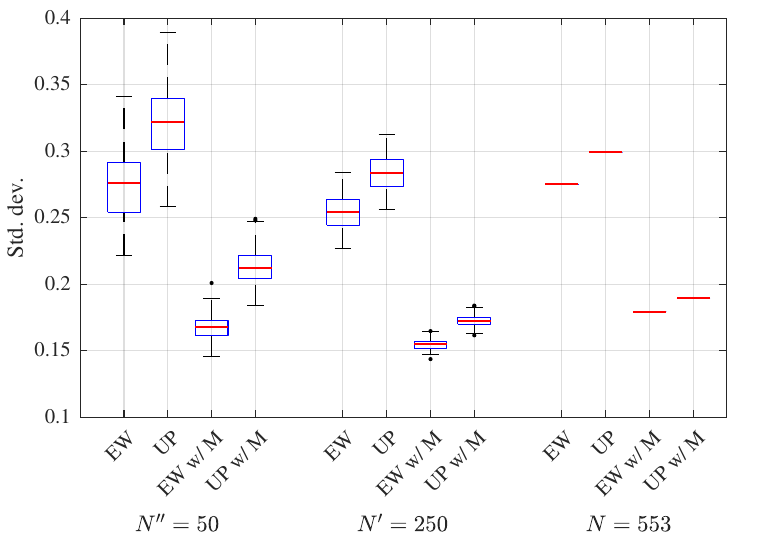}
\caption{Annualized standard deviations across bootstrap security subsets for the 4 different portfolio forming methods and the selected numbers of securities. Since the $N=553$ stocks case contains the full sample, the results for this case are single points rather than boxplots.}
\label{boxplot2}
\end{center}
\end{figure}

We also inspect portfolios formed using four different methods: (i) without the {momentum decorator} and using equal weights within the long and short parts of the portfolio (EW); (ii) without the {momentum decorator} and using utility proportional weighting (UP); (iii) with the {momentum decorator} and equal weighting (EW w/ M); (iv) with the {momentum decorator} and utility proportional weighting (UP w/ M). The annualized mean returns and standard deviations for the 4 different combinations across the bootstrap subsamples are displayed in Figures \ref{boxplot1} and \ref{boxplot2}.

\begin{figure}[H]
\begin{center}
\includegraphics[width=0.75\linewidth]{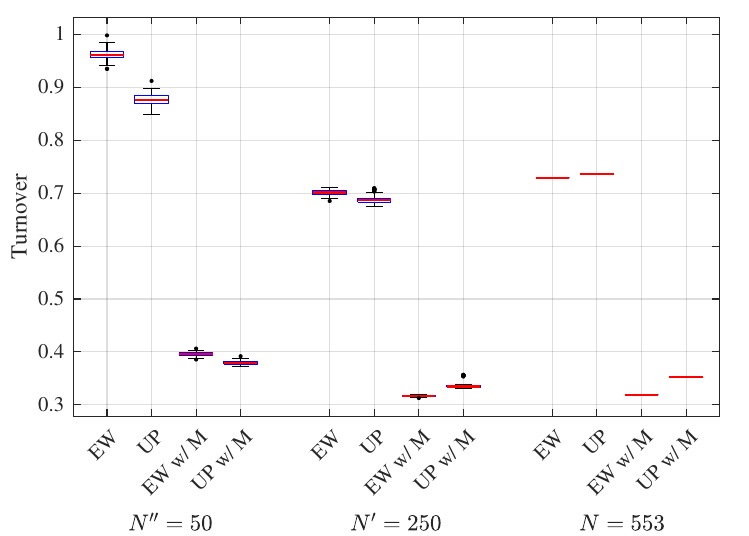}
\caption{Average daily turnover across bootstrap security subsets for the 4 different portfolio forming methods and the selected numbers of securities. Since the $N=553$ stocks case contains the full sample, the results for this case are single points rather than boxplots.}
\label{boxplot3}
\end{center}
\end{figure}

Several things can be noted from Figure \ref{boxplot1}. Firstly, the results in Table \ref{tbl:results1} are confirmed here as well - the annualized mean return is higher for the portfolios formed with more securities. In addition, portfolios with utility proportional weighting outperform those with the equal weighting scheme, in terms of annualized mean returns - for all the considered constellations. Moreover, the results also indicate that the inclusion of the {momentum decorator} improves the performance in terms of mean returns for the first two subsample sizes, but this is not the case for the full sample. However, as seen in Figure \ref{boxplot2} the {momentum decorator} reduces the portfolio risk as measured by standard deviation, in all considered cases. Moreover, as shown in Figure \ref{boxplot3}, the momentum decorator drastically reduces turnover in all examined scenarios, thus reducing the trading costs. The results suggest that, as expected, the proposed method does indeed thrive when larger sets of securities are considered, since generally more potentially profitable pairs will be present. Moreover, this effect of the asset universe size seems to most strongly affect the mean returns, with the momentum decorator and improved weighting mostly affecting the turnover -- which in turn does improve portfolio returns for the smallest portfolios, due to high transaction costs.

Finally, to inspect the ability of the proposed approach to generate portfolios robust to systematic risk sources with significant excess returns, we perform regression of portfolio returns on the Fama-French factors. We focus on the full security sample, and in addition to the long-short portfolio, we also include the long and short portfolios separately in order to study the origins of excess returns generated by the approach (shown in Table \ref{tbl:results2}).
\begin{table*}
\renewcommand{\arraystretch}{1.3}
\caption{Results of the regressions of the long, short and long-short strategy portfolio returns on the Fama-French 5 factors of the form: $R_t = \alpha_t + \beta_{MKT}MKT_t + \beta_{SMB}SMB_t + \beta_{HML}HML_t + \beta_{RMW}RMW_t + \beta_{CMA}CMA_t + \varepsilon_t$. Coefficients significant at the $0.05$ level are shown in boldface. The $R^2$ measure is presented in the adjusted form: $1-(1-R^2)(n-1)/(n-k-1)$, and the number of observations is $n=7037$.\vspace{10pt}}
\label{tbl:results2}
\centering
\begin{tabular}{llccccccc}%
    \toprule
    Method & Portfolio & $\alpha$ & $\beta_{MKT}$ & $\beta_{SMB}$ & $\beta_{HML}$ & $\beta_{RMW}$ & $\beta_{CMA}$ & $R^2$  \\ \hline\rule{0pt}{3ex}

    \multirow{3}{*}{EW}
    & Long       & $\phantom{-}\bm{0.060}$ & $\bm{1.137}$ & $\bm{0.548}$ & $\bm{0.511}$ & $\phantom{-}\bm{ 0.127}$ & $           \bm{-0.099}$ & $0.50$ \\
    & Short      & $    \phantom{-}0.005 $ & $\bm{0.879}$ & $\bm{0.518}$ & $\bm{0.308}$ & $\phantom{-}\bm{ 0.258}$ & $\phantom{-}\bm{ 0.226}$ & $0.55$ \\
    & Long-short & $\phantom{-}\bm{0.045}$ & $\bm{0.254}$ & $    0.030 $ & $\bm{0.204}$ & $           \bm{-0.134}$ & $           \bm{-0.329}$ & $0.05$ \\[6pt]
    \multirow{3}{*}{UP}
    & Long       & $\phantom{-}\bm{0.071}$ & $\bm{1.136}$ & $\bm{0.537}$ & $\bm{0.533}$ & $\phantom{-}\bm{ 0.093}$ & $                -0.100$ & $0.46$ \\
    & Short      & $              -0.002 $ & $\bm{0.878}$ & $\bm{0.516}$ & $\bm{0.299}$ & $\phantom{-}\bm{ 0.261}$ & $\phantom{-}\bm{ 0.226}$ & $0.49$ \\
    & Long-short & $\phantom{-}\bm{0.063}$ & $\bm{0.254}$ & $    0.022 $ & $\bm{0.237}$ & $           \bm{-0.170}$ & $           \bm{-0.331}$ & $0.04$ \\[6pt]
    \multirow{3}{*}{EW w/ M}
    & Long       & $\phantom{-}\bm{0.045}$ & $\bm{1.116}$ & $\bm{0.505}$ & $\bm{0.492}$ & $\phantom{-}\bm{ 0.165}$ & $           \bm{-0.064}$ & $0.73$ \\
    & Short      & $     \phantom{-}0.010$ & $\bm{0.859}$ & $\bm{0.475}$ & $\bm{0.295}$ & $\phantom{-}\bm{ 0.247}$ & $\phantom{-}\bm{ 0.196}$ & $0.71$ \\
    & Long-short & $     \phantom{-}0.023$ & $\bm{0.248}$ & $     0.025$ & $\bm{0.225}$ & $           \bm{-0.083}$ & $           \bm{-0.290}$ & $0.09$ \\[6pt]
    \multirow{3}{*}{UP w/ M}
    & Long       & $\phantom{-}\bm{0.049}$ & $\bm{1.123}$ & $\bm{0.503}$ & $\bm{0.487}$ & $\phantom{-}\bm{ 0.154}$ & $           \bm{-0.066}$ & $0.71$ \\
    & Short      & $               -0.001$ & $\bm{0.862}$ & $\bm{0.484}$ & $\bm{0.279}$ & $\phantom{-}\bm{ 0.246}$ & $\phantom{-}\bm{ 0.207}$ & $0.71$ \\
    & Long-short & $\phantom{-}\bm{0.039}$ & $\bm{0.260}$ & $     0.012$ & $\bm{0.208}$ & $           \bm{-0.084}$ & $           \bm{-0.279}$ & $0.09$ \vspace{3pt}\\\hline
\end{tabular}%
\end{table*}
These results indicate that, although the coefficients of the long-short portfolio regression on systematic risk factors seem to be statistically significant, the total portion of the explained variance, as measured by the adjusted $R^2$, is very low. This means that, even though the approach itself does not account for market beta and is not seeking explicitly market-neutral portfolios, the long-short portfolios have very little exposure to systematic risk factors including the market. The statistically significant alphas estimated in the regressions can be traced to two sources: firstly, the long-only portfolio representing the long part of the strategy seems to consistently outperform the market, as indicated by significant positive alphas; secondly, the long-only portfolio representing the short part of the strategy consistently underperforms. Both of these, since they are long-only portfolios, have relatively large systematic risk exposures, but the long-short combinations seem to mostly neutralize the systematic effects.

\section{Conclusion}\label{sec:conclusion}
In this paper, an approach for portfolio selection based on pairwise security preference relations using statistical arbitrage methods is proposed. The potential method is utilized to calculate the pairwise preferences function $\rho^*$ and form a preference relation graph $\mathcal{G}$ from the contradictory trading signals given by the pairwise preferences $\rho$ calculated from the data. The graph topology is used to select the most preferred securities to include in the portfolio, or to ultimately form a long-short portfolio. In addition, we propose a utility proportional weighting scheme within the long and short parts of the portfolios, and compare it to the equal-weighted approach. An important aspect of the proposed method is that it may be used in combination with any other pairs trading signal generation methods.

The proposed approach was tested on a dataset of U.S. stocks from 1990 to 2019, with a security bootstrap method to examine the robustness of the portfolios and study their performance with respect to the size of the asset universe. Even though we use the most basic pairs trading approach for calculating pairwise preference, without any complex trade timing algorithms, and assuming relatively high transaction costs in the backtest, the results show that the estimated portfolios perform well, especially when a large sample of securities is considered. These results justify the proposed methodology, and open the space for further possibilities in the context of more elaborate statistical arbitrage approaches. Moreover, we have addressed the problem of high turnover in large portfolios of securities with the use of the \emph{momentum decorator}, based on the security utilities from the potential method. This modification indeed leads to lower turnover by extending the holding period, which allows the position a more time to become profitable before exiting. In addition, the momentum decorator also reduced the volatility and increased robustness of security selection. A very favorable property of the proposed portfolio construction approach is the improved performance in the high-dimension regime, for which we provide theoretical and empirical evidence. In addition, the proposed method yields portfolios which are shown to be robust to systematic sources of risk, additionally affirming the proposed methodology. 


Some aspects of the potential method allow for further research and more elaborate strategies. Specifically, in the case of long-short portfolios it would be beneficial to explicitly seek market neutral portfolios rather than zero investment portfolios, since the former is not necessarily implied by the latter. The preference functions may be modified in order to include information about the systematic risk exposures of the securities. Moreover, we believe that including variance reduction techniques applied to the potential method is a potent avenue of further research which may improve the portfolio performance. The results shown in the paper will also hopefully inspire new research in terms of more advanced strategies and trade signal generating methods.

\section*{Funding}
This work was supported in part by the Croatian Science Foundation under Project 5241, and European Regional Development Fund under Grant KK.01.1.1.01.0009 (DATACROSS).

\bibliographystyle{apacite}

%
%
%
%
%

%
%
%
%
%
\pagebreak

\begin{appendices}
\numberwithin{equation}{section}
\numberwithin{figure}{section}

\section{Potential Method -- Derivation and Implementation Details}\label{sec:apx:potential_method_derivation} 

Let $\{s_1,\ldots, s_N\}\in\mathcal{S}$ be an indexed set of $N$ securities. The potential method defines a preference function $\rho^*$ as the one for which there exists a utility $u^*$ which satisfies the condition:
\begin{align}\label{eq:pm_equation}
    \rho^*(s_i, s_j) = u^*(s_i) - u^*(s_j)\quad \forall s_i, s_j \in \mathcal{S},\  i<j. 
\end{align}
\noindent It is clear that the preference function $\rho^*$ needs to be asymmetric, that is: $\rho(s_i, s_j) = -\rho(s_j, s_i)$. This implies that the preference function needs to be evaluated for $N \choose 2$ pairs (elements of $\mathcal{S}\times\mathcal{S}$) rather than $N^2$ pairs. The Equation \eqref{eq:pm_equation} can be written in matrix form as:
\begin{align}\label{eq:apx:potential_method_constraint}
    \bm{\rho}^* = \bm{B}\bm{u}^*
\end{align}
\noindent where $\bm{u}^*$ is the $N \times 1$ vector of security utilities, $\bm{\rho}^*$ is the ${N\choose 2} \times 1$ vector of pairwise preferences, and $\bm{B}$ is the ${N\choose 2} \times N$ graph incidence matrix.  
An example of the incidence matrix representing a (full) preference graph with $N=4$ securities looks like:
\newcommand{\bbinom}{\genfrac(){0pt}0}

\begin{align}
   \stackunder[8pt]{\begin{bmatrix}
\rho^*(s_1, s_2) \\
\rho^*(s_1, s_3) \\
\rho^*(s_1, s_4) \\
\rho^*(s_2, s_3) \\
\rho^*(s_2, s_4) \\
\rho^*(s_3, s_4) \\
\end{bmatrix}}{\bm{\rho}^*} = \stackunder[8pt]{\begin{bmatrix}
    1 &           -1 & \phantom{-}0 & \phantom{-}0 \\
    1 & \phantom{-}0 &           -1 & \phantom{-}0 \\
    1 & \phantom{-}0 & \phantom{-}0 &           -1 \\
    0 & \phantom{-}1 &           -1 & \phantom{-}0 \\
    0 & \phantom{-}1 & \phantom{-}0 &           -1 \\
    0 & \phantom{-}0 & \phantom{-}1 &           -1 \\
    \end{bmatrix}}{\bm{B}} \cdot \stackunder[2pt]{\begin{bmatrix}
u^*(s_1)\\
u^*(s_2)\\
u^*(s_3)\\
u^*(s_4)\\
\end{bmatrix}}{\bm{u}^*}
\end{align}

If there is no utility vector $\bm{u}^*$ for which the constraint (\ref{eq:apx:potential_method_constraint}) is satisfied given $\bm{\rho}$, then the function $\rho$ is not a preference function and an approximation of $\bm{\rho}$ needs to be found for which  (\ref{eq:apx:potential_method_constraint}) will hold for some $\bm{u}^*$. This task can be expressed as an optimization problem:
\begin{align}\label{eq:apx:potential_method_optimization}
    \bm{\rho}^* =& \argmin_{\bm{\rho}'\in\mathcal{P}} \quad d(\bm{\rho}', \bm{\rho})\\
    \textrm{s.t.}     \quad & \bm{B}\bm{u}^* = \bm{\rho}^* \nonumber \\
                            & \sum_{i=0}^N u_i^* = 0   \nonumber
\end{align}
\noindent where $\mathcal{P}$ is the set of all preference vectors $\bm{\rho}^*$ and $d$ is a distance function. The constraint $\sum_{i}^N u_i^* = 0$ is needed in order to obtain a unique solution, because a preference relation obtained from a utility function is invariant to scaling, i.e. $\forall {s_i,s_j}\in\mathcal{S},\,\forall\alpha\in\mathbb{R}^+\setminus\{0\}\ s_i \succ s_j \iff \alpha u^*(s_i) > \alpha u^*(s_j)$. In this paper, for the distance function we use the Euclidean distance:
\begin{equation}
    d(\bm{\rho}', \bm{\rho}) = ||\bm{\rho}' - \bm{\rho}||_2^2,
\end{equation}
for which an analytical solution exists\footnote{Other distances are also applicable, for example Manhattan distance, which would result in sparse residuals $\bm{r} = \bm{\rho}' - \bm{\rho}$. However, an analytical solution for the optimization problem (\ref{eq:apx:potential_method_optimization}) does not exist when using Manhattan distance and in our experiments we did not see significant improvements when substituting Euclidean for Manhattan distance.}. The optimization problem (\ref{eq:apx:potential_method_optimization}) can then be rewritten as:
\begin{align}
    \bm{u}^*   &= \argmin_{\bm{u}'} \left\{ ||\bm{Bu'} - \bm{\rho}||_2^2\right\}\\
    \textrm{s.t.} & \quad \sum_{i=0}^N u_i^* = 0 \nonumber
\end{align}
\noindent Note that $\bm{Bu'}$ can be substituted for $\bm{\rho}'$ in (\ref{eq:apx:potential_method_optimization}), since multiplying an incidence matrix with a utility vector will always yield a vector which corresponds to preference function vector $\bm{\rho}^*$. By taking the derivative w.r.t. $\bm{u'}$ and equating it with zero, the following system of equations is obtained:
\begin{align}
    \bm{B}^\tran\bm{B}\bm{u}^* &= \bm{B}^\tran\bm{\rho} \label{eq:apx:l2_norm_problem} \\
    \sum_{i=0}^N u_i^* &=  0.  \label{eq:apx:l2_norm_problem_constraint}
\end{align}
Equations (\ref{eq:apx:l2_norm_problem}) and (\ref{eq:apx:l2_norm_problem_constraint}) can be added together to get:
\begin{align}\label{eq:apx:l2_norm_final_problem}
    \left[\bm{B}^\tran\bm{B} + \bm{J}^\tran\right]\bm{u}^* = \bm{B}^\tran\bm{\rho},
\end{align}
where $\bm{J}$ is a matrix of ones with the same dimension as $\bm{B}^\tran\bm{B}$. Finally solving (\ref{eq:apx:l2_norm_final_problem}) for $\bm{u}^*$ results in:
\begin{align}
    \bm{u}^* = \left[ \bm{B}^\tran\bm{B} + \bm{J} \right]^{-1}\bm{B}^\tran\bm{\rho},
\end{align}
which can be further simplified with by employing the fact that $\bm{B}^\tran\bm{B}$ is the Laplacian of a complete graph which implies that term $\left[\bm{B}^\tran\bm{B} + \bm{J} \right]^{-1} = \frac{1}{N}\bm{I}$. Thus the final solution to the Equation (\ref{eq:apx:l2_norm_final_problem}) is:
\begin{align}\label{eq:apx:optimal_potentials}
    \bm{u}^* = \frac{1}{N}\bm{B}^\tran\bm{\rho}.
\end{align}
The preference vector $\bm{\rho}^*$ can be calculated by simply plugging $\bm{u}^*$ back into Equation (\ref{eq:apx:potential_method_constraint}):
\begin{align}
    \bm{\rho}^* = \bm{B}\bm{u}^*.
\end{align}

\section{Preference preserving transformations}\label{sec:preference_preserving_transformations}

A graph $\mathcal{G}^*$ is considered a preference graph if one can derive a preference relation $M^*$ from it, for example via the rule:
\begin{align}\label{eq:apx:potential_method_trading_signals_rule}
    M^*(s_i, s_j) = \begin{cases}
      sign(\rho^*(s_i, s_j))&\text{if }\rho^*(s_i, s_j) \neq 0\\
      0    & \text{otherwise.}
    \end{cases} 
\end{align} 
\noindent A preference preserving transformation is a transformation of the preference graph which results in a new preference graph. 
In order to show that a transformation of the preference graph is a preference preserving transformation we need to show that the relation $M$ derived from the transformed graph $\mathcal{G}^{*}$ satisfies the transitivity and irreflexivity properties (asymmetry is implied if transitivity and irreflexivity hold).

\subsection{Edge thresholding}
Thresholding transformation is defined as:
\begin{align}
    \rho^*_k(s_i, s_j) = \begin{cases}
      \rho^*(s_i,s_j)&\textrm{if } |\rho^*(s_i, s_j)| \geq k\\
      0&\textrm{otherwise}
    \end{cases}
\end{align}
By definition $\mathcal{G}^{*}$ does not have self referential links (e.g. $\rho^*(s_i, s_i)\neq 0$) and removing edges will not introduce new self referential links. Therefore, a relation $M$ derived from the preference graph transformed via thresholding $f(\mathcal{G}^*)$ will posses the irreflexivity property, since thresholding can only remove graph edges, not introduce them. Additionally, since the potential method will produce a graph with edge weights: $\rho(s_i, s_j) = u(s_i) - u(s_j)$, it can be shown that the transitivity property will also hold. 
Let $(s_i, s_j, s_k)\in\mathcal{S}$ be any triplet for which $M^*(s_i, s_j) = 1$, $M^*(s_j, s_k) = 1$, and $M^*(s_i, s_k) = 1$ holds. This is reflected in a preference graph as $\rho^*(s_i, s_j) > 0$, $\rho^*(s_j, s_k) > 0$, and $\rho^*(s_i, s_k) > 0$. In addition we know that $\rho^*(s_i, s_j) = u^*(s_i) - u^*(s_j)$, then:
\begin{align}
\rho^*(s_i, s_j) = u^*(s_i) - u^*(s_j) > 0 &\implies u^*(s_i) > u^*(s_j),\nonumber\\
\rho^*(s_j, s_k) = u^*(s_j) - u^*(s_k) > 0 &\implies u^*(s_j) > u^*(s_k),\nonumber\\
\rho^*(s_i, s_k) = u^*(s_i) - u^*(s_k) > 0 &\implies u^*(s_i) > u^*(s_k),\nonumber\\
u^*(s_i) > u^*(s_j) > u^*(s_k).&
\end{align}
Then we can derive  $\rho^*(s_i, s_k) > \rho^*(s_i, s_j)$ and $\rho^*(s_i, s_k) > \rho^*(s_j, s_k)$, by subtracting $u^*(s_k)$ and converting to $\rho^*$:
\begin{align}
    &u^*(s_i) > u^*(s_j) > u^*(s_k),\nonumber\\ 
    &u^*(s_i) - u^*(s_k) > u^*(s_j) - u^*(s_k) > u^*(s_k) - u^*(s_k),
    \nonumber\\
    &\rho^*(s_i,s_k) > \rho^*(s_j, s_k) > 0.
\end{align}
In analogy:
\begin{align}
    &u^*(s_i) > u^*(s_j) > u^*(s_k),\nonumber\\ 
    &u^*(s_i) - u^*(s_i) > u^*(s_j) - u^*(s_i) > u^*(s_k) - u^*(s_i),
    \nonumber\\
    &0 > \rho^*(s_j, s_i)  > \rho^*(s_k, s_i),
    \nonumber\\
    &0 > -\rho^*(s_i, s_j) > -\rho^*(s_i, s_k),
    \nonumber\\
    &\rho(s_i, s_k) > \rho^*(s_i, s_j) > 0.
\end{align}

\begin{figure*}[h]
    \centering%
    \includegraphics[width=\textwidth]{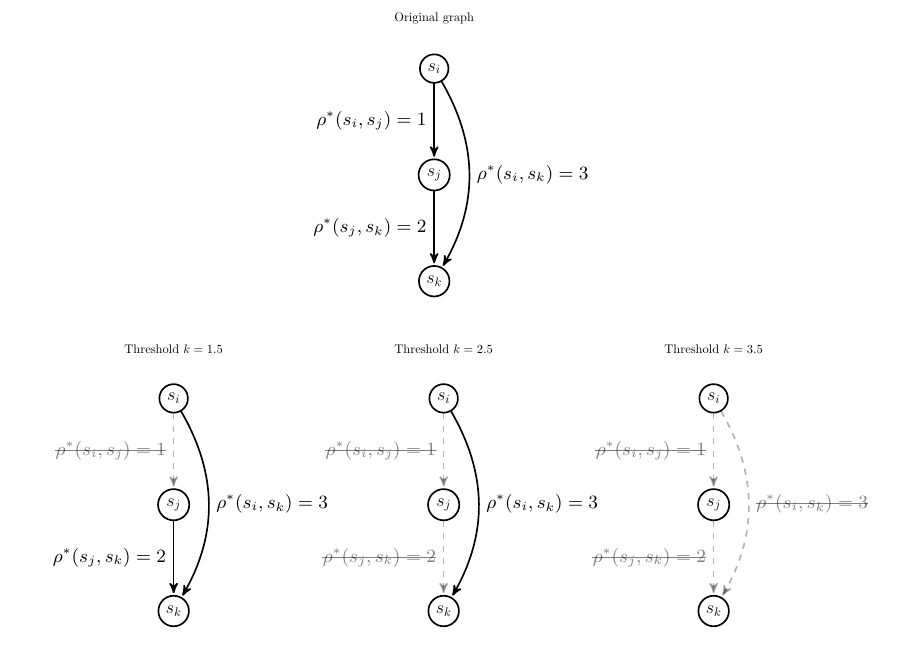}
    \caption{Preservation of transitivity property for thresholding transformation. Depending on the thresholding value $k$ some edges in the graph will be removed (bottom figures), however the transitivity property will never be violated.} 
    \label{fig:transitivity_property_preservation_thresholding}
\end{figure*}

Therefore $\rho^*(s_i, s_k) > \rho^*(s_i, s_j)$ and $\rho^*(s_i, s_k) > \rho^*(s_j, s_k)$ holds, which in turn means that by thresholding we will remove either $\rho^*(s_i, s_j)$ or $\rho^*(s_j, s_k)$ before removing $\rho^*(s_i, s_k)$. The resulting graph will not violate the transitivity property for the considered triplets and consequently the preference graph $\mathcal{G}^*$ itself. One such situation is shown in Figure \ref{fig:transitivity_property_preservation_thresholding}.

\subsection{Vertex pruning}

\begin{figure}[h!]
    \centering
    \includegraphics[width=0.58\textwidth]{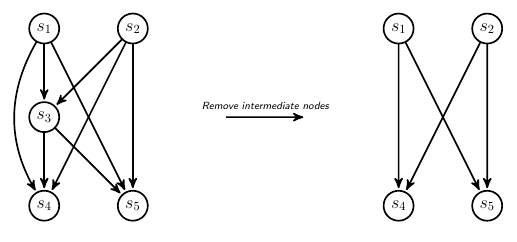}
    \caption{An example of removal of the intermediate vertex $s_3$, resulting in a bipartite graph. Since the maximum path length of the bipartite graph is $1$, the transitivity property is preserved.}
    \label{fig:connecting_most_preffered_and_least_preffered_nodes}
\end{figure}

The second transformation of the preference graph we utilize is removal of intermediate vertices - those which have both in- and out-degree higher than $0$. Applying such a transformation to the preference graph $\mathcal{G}^*$ result in a bipartite graph, which is a consequence of the transitivity property. A preference graph is a directed acyclic graph and as such it contains vertices with $in\_degree(s_i)=0 \land out\_degree(s_i) > 0$ and vertices with $in\_degree(s_j) > 0 \land out\_degree(s_j) = 0$, which we term \textit{sources} and \textit{sinks}, respectively. If a path from source vertex $s_i$ to sink vertex $s_j$ exists, then by transitivity property a path of length $1$ connecting $s_i$ to $s_j$ must also exist. Removal of intermediate nodes is equivalent to removal of all paths from source $s_i$ to sink $s_j$ with length exceeding $1$. After removing intermediate vertices, all paths will have length of $1$ and will connect source vertices to sink vertices, and the resulting graph will be a bipartite graph.
Bipartite graphs, by definition do not violate transitivity property since 
there are no triplets of vertices  $(s_i, s_j, s_k)$ for which following holds: $M^*(s_i, s_j) = 1 \land M^*(s_j, s_k) = 1$. As such there is no triplet $(s_i, s_j, s_k)$ for which the transitivity property could be violated. Irreflexivity is also preserved since no new edges are introduced. Therefore, removal of intermediate nodes in a preference graph $\mathcal{G}^*$ is also preference preserving graph transformation. This process can be seen in Figure \ref{fig:connecting_most_preffered_and_least_preffered_nodes}.

\subsection{Final vertex selection}
The final transformation of the preference graph used in this paper is ranking securities, i.e. vertices, w.r.t. utility $u^*$ and keeping only top $n$ and bottom $m$ vertices in the graph. In general this graph transformation is not a preference preserving transformation, however, when applied to the bipartite graph it is. Since the bipartite graph does not have any triplet $(s_i, s_j, s_k)$ for which the transitivity property could be violated, removal of vertices and its corresponding edges will preserve the preference structure of a graph. Since the operation is applied after removing intermediate nodes, i.e. constructing a bipartite graph, we ensure that resulting graph will still be a preference graph. 

\section{Potential method as an unbiased and consistent estimator of $u^*$ and $\rho^*$}\label{sec:apx:potential_method_estimator}

We here provide proof that, under some assumptions on the function $\rho$, the potential method is an unbiased and consistent estimator of utility $u^*$ and preference function $\rho^*$. In addition, we show that the variance of the potential method estimator diminishes with the number of considered securities. The potential method estimates the utility as:
\begin{align}
    \bm{u}^* &= \frac{1}{N}\bm{B}^\tran\bm{\rho} \label{eq:apx:utility_estimator}\\
    \text{s.t.} & \sum_{i} u^*(s_i) = 0. \nonumber 
\end{align}
\noindent Where $\bm{u}^*$ is the vector of estimated utilities and $\bm{\rho}$ is the vector of pairwise preferences. Let the pairwise preferences be generated by distorting the preference function $\rho^\dagger$ with zero-mean additive noise $\epsilon$, independent of the function $\rho$, as:
\begin{align}
    {\rho(s_i, s_j)}
    &= {\rho}^\dagger(s_i, s_j) + {\epsilon_{i,j}}. \label{eq:apx:preference_assumption_1}
\end{align} 
Using ${\rho}^\dagger(s_i, s_j) = u^\dagger(s_i) - u^\dagger(s_j)$ (where $u^\dagger$ is the true utility function), and $\sum_{j \neq i}u^\dagger(s_j) = - u^\dagger(s_i)$, the Equation (\ref{eq:apx:utility_estimator}) can be written in non matrix form as:

{
\begin{align}
    u^*(s_i) 
    &= \frac{1}{N} \left(\sum_{i < j} \rho(s_i, s_j) - \sum_{j < i} \rho(s_j, s_i) \right)\nonumber\\
    &= \frac{1}{N} 
        \sum_{j\neq i} \rho(s_i, s_j) 
    = \frac{1}{N}
    \sum_{j\neq i} \left( \rho^\dagger(s_i, s_j) + \epsilon_{i, j} \right) \nonumber\\
    &= \frac{1}{N}
    \sum_{j\neq i} \left( u^\dagger(s_i) - u^\dagger(s_j) + \epsilon_{i, j} \right) \nonumber\\
    &= \frac{1}{N}\left(
        (N-1)u^\dagger(s_i)  
        - \sum_{j \neq i} u^\dagger(s_j) +
        \sum_{j \neq i} \epsilon_{i,j}
    \right) \nonumber\\
    &= \frac{1}{N}\left(
        (N-1)u^\dagger(s_i) + u^\dagger(s_i) + \sum_{j \neq i} \epsilon_{i,j}
    \right) \nonumber\\
    &= u^\dagger(s_i) + \frac{1}{N} \sum_{j \neq i} \epsilon_{i,j} 
\end{align}
}%
\noindent Since $\E[{\epsilon_{i,j}}]=0\; \forall i,j$, the expected value of the estimated utility is now easily calculated as
:
\begin{align}
    \E[u^*(s_i)] 
    &= \E\left[u^\dagger(s_i) + \frac{1}{N} \sum_{j \neq i} \epsilon_{i,j}\right]\nonumber\\
    &= u^\dagger(x_i) + \frac{1}{N} \sum_{j \neq i}\E\left[ \epsilon_{i,j}\right]\nonumber\\
    &= u^\dagger(s_i),
\end{align}
showing that the potential method is an unbiased estimator of the utility $u^\dagger$ when the additive noise is zero-mean. The variance of the utility estimator can also be easily calculated as:
\begin{align}
    \Var&[u^*(s_i)] 
    = \Var\left[ \frac{1}{N} \sum_{j\neq i} \epsilon_{i,j}\right] \nonumber\\
    &= \frac{1}{N^2} \left( \sum_{j \neq i} \Var[\epsilon_{i,j}] + \sum_{j\neq i} \sum_{k \neq i\ \wedge\ k \neq j} \Cov[\epsilon_{i, j}, \epsilon_{i, k}] \right). 
\end{align}

If the noise terms $\epsilon_{i,j}$ are uncorrelated across pairs $i,j$ (i.e. zero covariance), with finite variance,
then with the increase in the number of considered securities $N$ the estimator $u^*$ converges in probability to $u^\dagger$, that is the potential method is a consistent estimator of utility ${u}^\dagger$. 
Moreover, if the noise terms all have equal variances $\sigma^2$, the variance of the estimator is then:
\begin{align}
    \Var[u^*(s_i)] = \frac{(N-1)}{N^2}\sigma^2.
\end{align}

\begin{figure}[h!]
    \centering
    \includegraphics[width=0.95\linewidth]{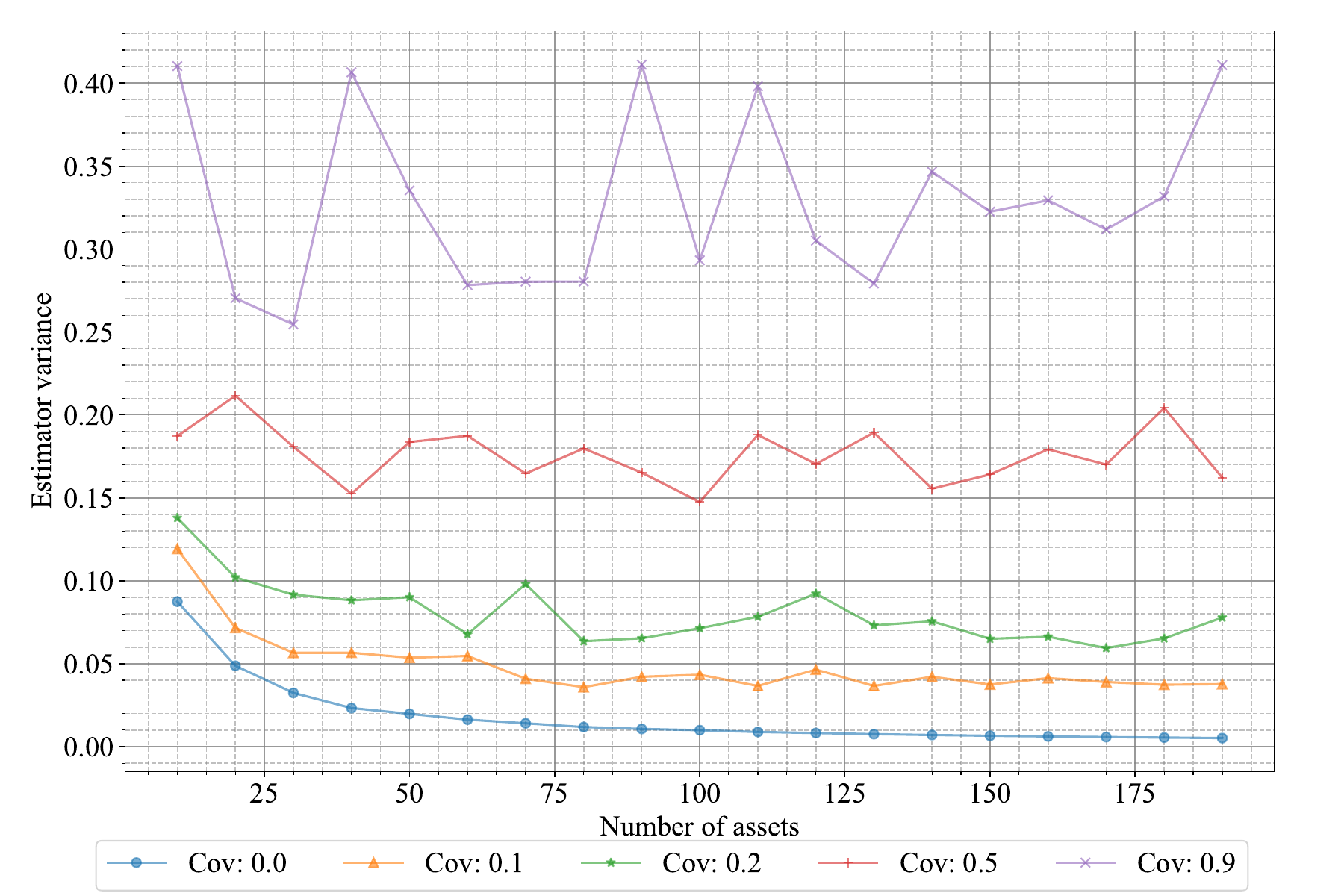}
    \caption{The variance of the potential method estimator of utility (y axis) for simulated data with $N$ assets (x axis). The noise terms ${\epsilon_{i,j}}$ were generated using a zero-mean Gaussian distribution with variance $\sigma^2=1$ and the same covariance across all pairs. The considered cases for covariance of noise terms include: $0.0, 0.1, 0.2, 0.5,$ and $0.9$, marked in different colors.}
    \label{fig:cov-var-dependence}
\end{figure}

\noindent The effects of noise covariance and asset dimensionality $N$ can be seen in the Figure \ref{fig:cov-var-dependence}. 
In the case of uncorrelated noise terms, the variance of the estimator is reduced towards $0$ as expected. However, when the noise terms are correlated the increase in the number of securities will not reduce the variance of the estimator to the zero, even in the limit. 

It is also easy to show that the potential method is an unbiased estimator of the preference function $\rho^*$. Since $\rho^*(s_i, s_j) = u^*(s_i) - u^*(s_j)$, then:
\begin{align}
    &\E[\rho^*(s_i, s_j)] 
    = \E[u^*(s_i) - u^*(s_j)]\nonumber\\
    &= \E\left[u^\dagger(s_i) - u^\dagger(s_j) + \frac{1}{N}\sum_{k\neq i}\epsilon_{i, k} - \frac{1}{N}\sum_{l\neq j}\epsilon_{j, l}\right]\nonumber\\
    &= u^\dagger(s_i) - u^\dagger(s_j) + \frac{1}{N}\sum_{k\neq i}\E\left[\epsilon_{i, k}\right] - \frac{1}{N}\sum_{l\neq j}\E\left[\epsilon_{j, l}\right]\nonumber\\
    &= u^\dagger(s_i) - u^\dagger(s_j)\nonumber\\
    &= \rho^\dagger(s_i, s_j).
\end{align}

The variance of the preference function can also be easily derived as:
\begin{align}
    &\Var\left[\rho^*(s_i, s_j)\right] 
    = \Var\left[u^*(s_i) - u^*(s_j)\right] \nonumber\\
    &= \Var\left[u^\dagger(s_i) + \dfrac{1}{N}\sum_{k\neq i}\epsilon_{i, k} - u^\dagger(s_j) - \frac{1}{N}\sum_{l\neq j} \epsilon_{j, l}\right]\nonumber\\
    &=\frac{1}{N^2} \Var\left[\sum_{k\neq i}\epsilon_{i,k} - \sum_{l\neq j}\epsilon_{j, l}  
    \right]\nonumber\\
    &=\frac{1}{N^2}
    \Bigg(
    \sum_{k\neq i}\Var\left[\epsilon_{i,k}\right] + \sum_{l\neq j}\Var[\epsilon_{j,l}]
    - \sum_{k \neq i} \sum_{l \neq j} \Cov[\epsilon_{i, k}, \epsilon_{j, l}]\nonumber\\
    &\quad + \sum_{k\neq i}\sum_{l \neq i \land l \neq k}\Cov[\epsilon_{i, k}, \epsilon_{i, l}]
    + \sum_{k\neq j}\sum_{l \neq j \land l \neq k}\Cov[\epsilon_{j, k}, \epsilon_{j, l}]
    \Bigg).
\end{align}
\noindent If the noise terms $\epsilon_{i,j}$ are uncorrelated across pairs $i,j$ (i.e. zero covariance), with finite variance,
then the potential method is a consistent estimator of the preference function, and it decreases with growing $N$. Moreover, in analogy with the utility estimator, if the variances of the noise terms are all $\sigma^2$, the variance of the preference function estimator is:

\begin{align}
    \Var[\rho^*(s_i, s_j)]
    &= \frac{1}{N^2}\left(\sum_{k\neq i}\Var[\epsilon_{i, k}] + \sum_{l\neq j}\Var[\epsilon_{j, l}]\right)\nonumber\\
    &= \frac{1}{N^2}\left(\sum_{k\neq i}\sigma^2 + \sum_{l\neq j}\sigma^2\right)\nonumber\\
    &= \frac{2(N-1)}{N^2}\sigma^2.
\end{align}

\end{appendices}

\end{document}